\documentclass[prb,twocolumn,showpacs,superscriptaddress,preprintnumbers,amssymb]{revtex4}
\usepackage{graphicx}% Include figure files
\usepackage{dcolumn}% Align table columns on decimal point
\usepackage{bm}% bold math
\usepackage{amssymb}
\usepackage{amsfonts}
\usepackage{amsmath}
\begin{document}
\title{Optimal thermoelectric figure of merit of
a molecular junction}
\author{Padraig Murphy}
\affiliation{Department of Physics, University of California,
Berkeley, CA 94720}
\author{Subroto Mukerjee}
\affiliation{Department of Physics, University of California,
Berkeley, CA 94720} \affiliation{Materials Sciences Division,
Lawrence Berkeley National Laboratory, Berkeley, CA 94720}
\author{Joel Moore}
\affiliation{Department of Physics, University of California,
Berkeley, CA 94720} \affiliation{Materials Sciences Division,
Lawrence Berkeley National Laboratory, Berkeley, CA 94720}
\date{\today}
\begin{abstract}
We show that a molecular junction can give large values of the
thermoelectric figure of merit $ZT$, and so could be used as a
solid state energy conversion device that operates close to the
Carnot efficiency. The mechanism is similar to the Mahan-Sofo
model for bulk thermoelectrics --- the Lorenz number goes to zero
violating the Wiedemann-Franz law while the thermopower remains
non-zero. The molecular state through which charge is transported
must be weakly coupled to the leads, and the energy level of the
state must be of order $k_B T$ away from the Fermi energy of the
leads. In practice, the figure of merit is limited by the phonon
thermal conductance; we show that the largest possible
$ZT\sim(\tilde{G}_{th}^{ph})^{-1/2}$, where $\tilde{G}_{th}^{ph}$
is the phonon thermal conductance divided by the thermal
conductance quantum.
\end{abstract}


\maketitle
\def\beq{\begin{equation}}
\def\eeq{\end{equation}}

%\section{Introduction}
%\label{sec:intro}

%Thermoelectrics: \cite{mahan_phys_today},\cite{disalvo_review},\cite{majumdar_review},\cite{mahan_sofo, humphrey}; \cite{mukerjee_moore};\cite{peterson_2007};
%$U$ in benzene: \cite{kivelson}, \cite{gammel}, \cite{baeriswyl},
%\cite{kivelson_et_al_reply};
%Molecule experiments: \cite{quek};
%Thermal conductance of self-assembled monolayers:\cite{wang_segalman_majumdar}.
%Thermal transport in molecules (Nitzan)\cite{nitzan_2007}, \cite{galperin_2007}.
%Transport: \cite{ng_lee}, \cite{meir_wingreen_1992},
%\cite{meir_wingreen_lee-1993}, \cite{meir_wingreen_lee-1991}.

%\cite{mehta},
%Kondo screening cloud: \cite{affleck_simon}, \cite{hand}, \cite{sorensen_affleck},
%Failures of DFT for nanoscale transport: \cite{wang}, \cite{koentopp}, \cite{toher},
%Kondo thermopower: \cite{dong_lei}, \cite{kim_hershfield}, \cite{scheibner-2005},
%Polygamma function: \cite{dingle}

Electrical transport in mesoscopic and nanoscale systems has been a major
focus of recent work in physics, chemistry, and materials science.  This interest has been motivated both by the potential for applications in electronics and by the opportunity to study novel transport phenomena such as conductance quantization~\cite{vanwees} and Coulomb blockade.
%, and the plateau in the electrical conductance associated with the Kondo effect \cite{ng_lee,meir_wingreen_lee}.
%\cite{ng_meir_wingreen_lee}.
Compared to electrical transport, thermal transport has been much
less intensively studied, despite a number of fascinating
results including the observation in 2000 of the quantum of thermal
conductance~\cite{schwab,rego_1997}.
%\cite{schwab,chiatti,rego_1997}.
%, and its universality among fermions, bosons and anyons\cite{rego_1999}.
Thermal and thermoelectric transport are also relevant to technological questions of great
importance, such as the construction of solid-state energy-conversion
devices.

The efficiency of such devices is determined by the
thermoelectric figure of merit~\cite{mahan_phys_today} $ZT$,
defined as
\[
ZT = \frac{G S^2}{ G^{el}_{th}/T + G^{ph}_{th}/T}.
\]
Here $G$ is the charge conductance, $S$ the thermopower,
$G^{el}_{th}$ the electron thermal conductance, and
$G^{ph}_{th}$ the phonon thermal conductance.
As $ZT\rightarrow\infty$, the device attains the Carnot efficiency.
The figure of merit of the
semiconductors typically used for thermoelectric applications
is approximately 1; $ZT$ values of $2$ or $3$ would
lead to a considerable increase in commercial utility.
Increased interest in $ZT$ has given additional motivation to work on
the thermal and thermoelectric properties of unusual materials:
nanowires~\cite{hicks_dresselhaus_nanowires,murphy_moore},
nanostructured materials~\cite{harman}, strongly correlated
materials~\cite{terasaki_1997,cavaong,mukerjee_moore}, and molecular
junctions~\cite{selzer_allara,pramod}. Of these, molecular
junctions are particularly promising, as it is hoped that the
phonon thermal conductance will be small due to a density of
states mismatch between the phonons in the leads and the phonon
modes of the molecule.
%Unfortunately, experimental work on even the electrical conductance has proven difficult, primarily due to metal-molecule contacts.
Improvements of standard density functional methods are needed to give unambiguous predictions
for the transport properties of molecular systems~\cite{quek,toher}, especially when
correlations between electrons are significant.
It is therefore valuable to have a thorough understanding of transport in
simplified models, such as the Anderson model (reviewed below), that describe
intramolecular correlations, both as a check on the accuracy of new {\it ab initio} methods
and to guide both experiment and numerics to where high $ZT$ may occur.

In this paper we find the figure of merit of a molecular junction. We start from the
Anderson model for the molecule and leads, which is valid under the
assumption that a single molecular level dominates transport.  If the
lifetime $\tau$ of an electron in the molecular state satisfies $\tau
\ll \hbar/(k_BT)$, it is generally true~\cite{sakano,unpub} that $ZT \ll
1$. We concentrate therefore on the regime $\tau \gg \hbar/(k_BT)$.
Physically, this corresponds to a wavefunction for the level that is
strongly localized at the center of the molecule, so that the state is
weakly coupled to the leads.  In reference (\onlinecite{tsaousidou}) it was
shown that for the Anderson model in this limit of weak coupling to the
leads, with strong interactions, and ignoring the phonon contribution to
the thermal conductance, the figure of merit becomes infinite. Here we
find that the phenomenon is surprisingly general ---  it is also true in
the limit of weak electron-electron interactions. We clarify the
mechanism in both cases by comparing with the work of Mahan and Sofo
\cite{mahan_sofo} on bulk thermoelectrics. (The mechanism, a violation
of the Wiedemann-Franz law, has previously been studied in quantum dots
in different parameter regimes in references (\onlinecite{boese_fazio,krawiec,kubala}).)
Also, by carrying the calculation for the electron-only $ZT$ to higher
order in the coupling to the leads, we find the optimal value of $ZT$ once
the phonon contribution is included. Finally, we suggest a candidate
molecule where high $ZT$ might be observed.

The Anderson model is defined by the Hamiltonian
\begin{align}
\label{eqn:anderson}
H &= -t\sum_{j>0, j<-1, \sigma}( {c^{\dagger}}_{j+1, \sigma} c_{j, \sigma} + hc)
\nonumber \\
&-\sum_{\sigma}\left\{t'_R({c^{\dagger}}_{1,\sigma} c_{0,\sigma} + hc) +
t'_L({c^{\dagger}}_{0,\sigma} c_{-1,\sigma} + hc)\right\}\\
&+ \varepsilon_d (n_{0\uparrow} + n_{0\downarrow} ) +
 Un_{0\uparrow}  n_{0\downarrow} .\nonumber
\end{align}
The label $\sigma$ represents the spin of the electrons; $j$ is the site index,
with the molecule represented by the site at $j=0$. The amplitude
to hop from the molecule to the right (left) lead is given by $t'_{R(L)}$, and the leads
 are represented by the sites
at $j>0$ and $j<0$. The energy to place a single electron on the
molecule is $\varepsilon_d$, which can be interpreted as a gate
voltage. The electron-electron repulsion within the molecule
results in an additional energy cost of $U$ to add two electrons to the
molecular site. The hopping within the leads is $t$, which gives a
band structure
%\cite{ashcroft_and_mermin}
of
$E(k) = - 2t\cos ka $, where $a$ is the lattice spacing.

As a first step, we find the transport properties with no
electronic repulsion ($U=0$), when the transmission function of the impurity
can be found by solving the single-particle Schr\"odinger
equation. The result is
\[
{\cal T}_{\sigma} (E) = \frac{\Gamma_L\Gamma_R(1-E^2/(4t^2)) }{
\overline\Gamma^2(1-E^2/(4t^2)) + (E(1-\overline\Gamma/(2t))-\varepsilon_d)^2},
\]
where $E$ is the energy of the incident electron, $\Gamma_{L(R)} = 2{t'}_{L(R)}^{2}/t$,
and $\overline\Gamma = (\Gamma_L+\Gamma_R)/2$.
For electrons close to the center of the band, and $\Gamma_{L, R} \ll t$,
we get the familiar Lorentzian transmission
\[
{\cal T}_{\sigma} (E) = \frac{\Gamma_L\Gamma_R }{ \overline\Gamma} 
\frac{\overline\Gamma }{
\overline\Gamma^2 + (E-\varepsilon_d)^2} .
\]
The hybridization energy, $\overline\Gamma$, determines the mean
lifetime, $\tau$, of an electron on the dot, through $\tau =
\hbar/\overline\Gamma$. We will assume this form of ${\cal
T}_\sigma(E)$ in the study of the non-interacting case that
follows.

Within the Landauer formalism the transmission function determines
the charge current $I$ and the heat current due to the electrons
$I^{el}_Q$ through the relations
\begin{align*}
I &= -\frac{e}{ h} \int_{-2t}^{2t} dE
({\cal T}_\uparrow (E)+{\cal T}_\downarrow (E)) (f_L^0(E) - f_R^0(E) ), \\
I^{el}_Q &= \frac{1}{ h} \int_{-2t}^{2t} dE (E-\mu)
({\cal T}_\uparrow (E)+{\cal T}_\downarrow (E))
(f_L^0(E) - f_R^0(E) ).
\end{align*}
Here $-e$ is the electron charge, and the
function $f_L^0$ is defined by  $f_L^0(E) = (e^{(E-\mu_L)/(k_B
T_L)} + 1)^{-1}$, where $\mu_L$ and $T_L$ are the chemical
potential and temperature in the left lead (and similarly for
$f_R^0$). At linear response, and for a chemical potential away
from the band edges, we have
\begin{align}
\label{eqn:curr1}
I &= -\frac{2e }{ h}\frac{\gamma_L\gamma_R }{\overline\gamma}
\left({\cal F}_0(\overline\gamma, \delta) eV + {\cal F}_1(\overline\gamma, \delta)  k_B \Delta T\right),\\
\label{eqn:curr2}
I^{el}_Q & = \frac{2 k_B T }{ h}\frac{\gamma_L\gamma_R }{\overline\gamma}
\left( {\cal F}_1(\overline\gamma, \delta)  eV + {\cal F}_2(\overline\gamma, \delta)  k_B \Delta T \right),
\end{align}
where $\gamma_{L(R)} = \Gamma_{L(R)}/(k_BT)$, $\overline\gamma = \overline\Gamma/(k_BT)$, $\delta = (\varepsilon_d-\mu)/(k_BT)$,
$eV=\mu_L-\mu_R$, and $\Delta T=T_L-T_R$.
The functions ${\cal F}_n$ can be found from equations (\ref{eqn:curr1}) and (\ref{eqn:curr2}) by setting
$x = (E-\mu)/(k_BT)$; the result is
\[
{\cal F}_n (\overline\gamma, \delta) =
\int_{-\infty}^\infty dx \frac{x^n}{ (2\cosh(x/2))^2}
\frac{\overline\gamma }{ \overline\gamma^2 + (x-\delta)^2}.
\]
${\cal F}_0$, ${\cal F}_1$ and ${\cal F}_2$ can then be expressed as
\begin{align*}
{\cal F}_0 & = \frac{1}{ 4\pi}
\left\{ \psi \left( \frac{\pi + w}{ 2\pi} \right)
+\psi \left( \frac{\pi + w^*}{ 2\pi} \right)\right\},\\
{\cal F}_1 & = \frac{1 }{ 4\pi i}
\left\{ w\psi \left( \frac{\pi + w}{ 2\pi} \right)-w^*
\psi \left( \frac{\pi + w^*}{ 2\pi} \right)\right\},\\
{\cal F}_2  &= \overline\gamma -\frac{1}{ 4\pi}
\left\{
w^2\psi \left( \frac{\pi + w}{ 2\pi} \right)
+w^{*2}\psi \left( \frac{\pi + w^*}{ 2\pi} \right)
\right\},
\end{align*}
where $w = \overline\gamma +i\delta$. The function $\psi$ is the
trigamma function $\psi(z) = \sum_{n=0}^\infty  (z+n)^{-2}$.

\begin{figure}
\includegraphics[width=3in,height=2.2in]{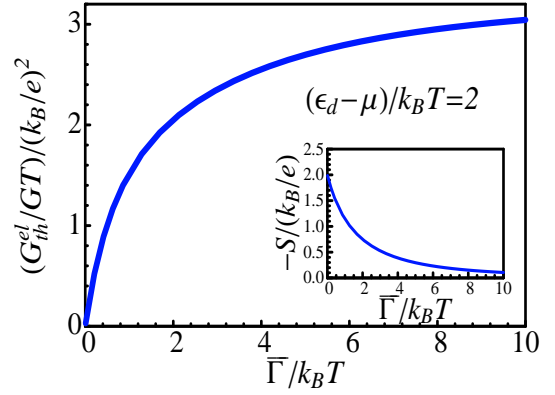}
\caption{The Lorenz number (main figure) and the thermopower (inset) plotted
as a function of $\overline\Gamma/(k_BT)$, with $(\varepsilon_d-\mu)/(k_BT)=2$ in both cases.
Only the electron contribution to the thermal conductance is included in the
calculation of the Lorenz number.
}
\label{fig:ZTofdelt_U=0}
\end{figure}

We define the limit of a long-lived molecular state
as $\overline\Gamma \ll k_B T$, i.e.~$\overline\gamma \ll 1$.
Ignoring the phonon contribution
to the heat current (to be restored below),
the figure of merit is given by
$ZT = \left({{\cal F}_0  {\cal F}_2/{\cal F}_1^2} - 1\right)^{-1}$.
%and is plotted in Fig.~\ref{fig:ZTofG_U=0}.
For $\overline\gamma$ fixed, $ZT$ is largest when
$\delta \simeq \pm 2.4$. If one holds
$\delta$ fixed, and allows $\overline\gamma\rightarrow 0$,
the figure of merit
diverges as
\[
ZT= \frac{1}{ \overline\gamma}\frac{\pi \delta^2}{ (2\cosh(\delta/2))^2} +
{\cal O}(\overline\gamma^0).
\]
(This follows from expanding ${\cal F}_n$ to first order in
$\overline\gamma$.)
As shown in Fig.~\ref{fig:ZTofdelt_U=0}, the
thermopower is finite as $\overline\gamma\rightarrow 0$,
and the divergence in $ZT$ is due to a violation of the Wiedemann-Franz law,
with the thermal conductance
vanishing faster than the electrical
conductance.

%\begin{figure}
%\includegraphics[width=3in,height=2.2in]{opt515_gr23.eps}
%\caption{The figure of merit, $ZT$, in the non-interacting case $(U=0)$, and
%without including the phonon contribution to the thermal conductance. In the main
%figure, it is shown as a function of $\overline\gamma = \overline\Gamma/(k_BT)$,
%with $\delta=(\varepsilon_d-\mu)/(k_BT)=2$;
%in the inset, $\overline\gamma=0.1$, and $ZT$ is plotted as a function of
%$\delta$.
%}
%\label{fig:ZTofG_U=0}
%\end{figure}

In practice, the figure of merit of the junction will be limited by the
phonon thermal conductance. This can be parametrized in terms of the fraction
$\tilde{G}_{th}^{ph}$
of a thermal conductance quantum transmitted by the molecule, i.e.
$
G_{th}^{ph} = \tilde{G}_{th}^{ph} \frac{\pi^2 }{ 3} \frac{k_B^2}{ h} T.
$
Using
\[
GS^2 = \frac{\gamma_L\gamma_R }{ \overline\gamma} \frac{2k_B^2}{ h} 
\frac{\pi \delta^2}{ (2\cosh(\delta/2))^2},
\]
%(this can be derived by expanding ${\cal F}_n$ to zeroth order in $\overline\gamma$)
we include the phonon contribution to get
%\[
%{1\over ZT}
%= {G_{th}^{e}/T \over G S^2} + {G_{th}^{ph}/T \over G S^2}
%             \simeq
%{\overline\gamma\over 1.32}  + {1.25\tilde{G}_{th}^{ph} \over\overline\gamma}
%{\overline\gamma^2\over\gamma_L\gamma_R}.
%\]
%($\Pi = GS^2$ is plotted in Fig.~\ref{fig:PFofG_U=0}.)
\[
\frac{1}{ZT}
= \frac{G_{th}^{e} }{ G S^2T} + \frac{G_{th}^{ph} }{ G S^2T}
             \simeq
\frac{(2\cosh(\delta/2))^2 }{ \pi \delta^2} \left( \overline\gamma
+ \frac{\pi^2 \tilde{G}_{th}^{ph}\overline\gamma }{ 6 \gamma_L\gamma_R}
\right).
\]
Assuming $\gamma_L=\gamma_R=\gamma$ and minimising $1/ZT$ with respect to $\gamma$
and $\delta$
gives, for small $\tilde{G}_{th}^{ph}$, an optimal value of
\[
ZT \simeq 0.51/(\tilde{G}_{th}^{ph})^{1/2}.
\]
The figure of merit becomes infinite as $\tilde{G}_{th}^{ph}$ goes to zero, as we found
before.
\begin{figure}
\includegraphics[width=3in,height=2.2in]{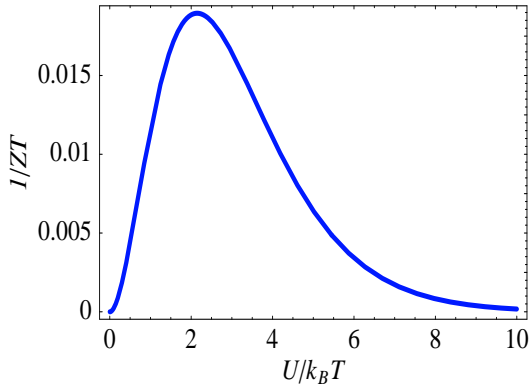}
\caption{
$1/ZT$, without the phonon contribution to the thermal conductance,
 is plotted for $\Gamma/(k_BT) \rightarrow 0$ and $(\varepsilon_d-\mu)/(k_B T)=2$.
}
\label{fig:ZTU}
\end{figure}

There is a simple physical picture that gives us the correct expressions for
${\cal F}_0$, ${\cal F}_1$, and ${\cal F}_2$ to order $\overline\gamma^0$,
and thus gives $1/ZT = 0 + {\cal O}(\overline\gamma^0)$.
This approximation
is known as ``sequential tunneling'', and has been used to study transport in
quantum dots and molecules \cite{beenakker-1991,beenakker_staring,koch,zianni}.
Suppose the molecular level has energy $\varepsilon_d$. Then, in this approximation,
an electron in the leads can tunnel onto the level only if it too has energy
$\varepsilon_d$. Energy is therefore conserved throughout the transport process.
This allows for a current proportional to $\Gamma_L\Gamma_R/\overline\Gamma$. In
quantum-mechanical tunneling
by contrast, the intermediate state is virtual; away from
resonance,  the current is
proportional to the probability to tunnel onto the level multiplied by
the probability to tunnel off,
and thus to $\Gamma_L\Gamma_R$.

In the sequential tunneling limit, each electron passing
through the junction carries the same amount of heat, $Q =
\varepsilon_d-\mu$. To understand why the Wiedemann-Franz law is
violated, it is essential to bear in mind that the thermal
conductance is defined as the heat current divided by the
temperature difference {\it under the condition of zero charge
current}. The condition $I=0$ implies that the flux of left-moving
electrons is the same as the flux of right-moving electrons; since
each left or right moving electrons carries exactly the same
amount of heat $Q$, the thermal current $I^{el}_Q$ is also zero.
This is the same mechanism that Mahan and Sofo used to model the
bulk material with the best thermoelectric figure of
merit~\cite{mahan_sofo,humphrey}.  For long-lived
molecular states, unlike bulk materials (except in the atomic limit~\cite{mukerjee_moore}), the theoretical analysis
can be extended to include electron-electron repulsion, as now
done.

We now consider whether, once electron-electron repulsion
in the molecule is added,
the electronic figure of merit $ZT$ still scales like $1/\overline\gamma$ in the limit of
a long-lived molecular state.
Define the probability of the molecular level being empty as $P(0)$,
of there being an up spin as $P(\uparrow)$, a down spin as $P(\downarrow)$,
and of it being doubly occupied as $P(2)$.  We consider the
currents where the molecule meets the left lead.
An example of a process that contributes to the currents at this point
is the tunneling of an up electron with energy $\varepsilon_d$ from the left
lead onto the empty level.
%The contribution of this process to the currents is
%\begin{align*}
%I &= \cdots-e \tau^{-1} P(0) f_{L\uparrow}^0 (\varepsilon_d) +\cdots,\\
%I^{el}_Q &= \cdots (\varepsilon_d-\mu)
%\tau^{-1} P(0) f_{L\uparrow}^0 (\varepsilon_d) +\cdots.
%\end{align*}
To find the total currents, we consider all possible tunneling processes, both onto and
off of the molecule.
%We assume that
%there is no magnetic field, which implies that
%$f_{L\uparrow}^0 = f_{L\downarrow}^0=f_{L}^0$, and
%similarly for the right lead.
For
$\delta>0$, ie. $\varepsilon_d>\mu$,
the charge and heat currents at the left junction
are then given by
\begin{align*}
I &= -e\tau_L^{-1} \left( P(0) 2 f_L(\varepsilon_d) -P(1)(1 - f_L(\varepsilon_d) ) \right. \\
 &\left.\qquad +P(1)f_L(\varepsilon_d+U)  - 2P(2) (1 - f_L(\varepsilon_d+U))\right),\\
I^{el}_Q &= \tau_L^{-1} k_B T \left( \delta P(0) 2 f_L(\varepsilon_d) -\delta P(1)(1 - f_L(\varepsilon_d) ) \right. \\
& \qquad +(\delta+u)P(1)f_L(\varepsilon_d+U)  \\
&\left.\qquad - (\delta+u)2P(2) (1 - f_L(\varepsilon_d+U))\right),
\end{align*}
where $P(1) = P(\uparrow)+P(\downarrow)$,
$u= U/(k_B T)$, and, as before, $\delta = (\varepsilon_d-\mu)/(k_BT)$.
The probabilities $P(0)$, $P(\uparrow)$, $P(\downarrow)$, and $P(2)$ are determined
from the steady state condition.
The final result for the linear response transport coefficients is
\begin{align*}
G &=
\frac{e^2}{ \hbar}\frac{\gamma_L \gamma_R }{\overline\gamma}\frac{e^{\delta +u} ( 1 + e^\delta(2+e^{\delta+u}))}{ (1+ e^{\delta} )(1+e^{\delta +u} )( 1 + e^{\delta+u}(2+e^{\delta}))},\\
S &= -\frac{k_B}{e} \frac{(\delta (1 +e^\delta(2+e^{\delta+u})) + u(1+e^\delta))
}{
( 1 + e^\delta(2+e^{\delta+u}))},\\
\frac{G_{th}^{el} }{ T} &= \frac{k_B^2}{ \hbar} \frac{\gamma_L \gamma_R }{\overline\gamma}
\frac{u^2 e^{2\delta + u} }{ (1+2e^{\delta+u} + e^{2\delta +u})
(1 + 2e^\delta + e^{2\delta +u}) } .
\end{align*}
(It is straight-forward to confirm that this result, with $u=0$, agrees with the
exact result for the non-interacting case in the limit of $\gamma_{L,R}\ll 1$, if
one makes the identification $\tau_{L(R)}^{-1} = \Gamma_{L(R)}/\hbar$.
This confirms that only sequential tunneling is important in the limit
of a long-lived state.)

%Setting $u=0$, the conductance is
%We confirm that this model, with $u=0$, agrees with the exact result for the non-interacting case in the limit of $\gamma\ll 1$.
%Setting $u=0$, the conductance is
%\[
%G = {e^2 \tau^{-1} \over k_B T} {e^\delta (1+e^\delta)^2 \over
%(1+e^\delta)^4} = {e^2 \tau^{-1} \over k_B T} {1\over (2\cosh(\delta/2))^2}.
%\]
%Identifying $\tau^{-1} = \Gamma/\hbar$ gives
%\[
%G = \gamma {e^2 \over h/(2\pi)}{1\over (2\cosh(\delta/2))^2}
% = {2e^2 \over h} {\gamma \pi \over (2\cosh(\delta/2))^2},
%\]
%which is the correct result.
%{\bf INCLUDE $\kappa$, $S$}

%Returning to the interacting case, the figure of merit is given by

%\lookhere
The figure of merit in the sequential tunneling
approximation (again ignoring phonons) is given by
\[
ZT = \frac{e^{-\delta} }{ u^2} \frac{ (\delta(1+2e^{\delta} + e^{2\delta+u}) + u(1+e^\delta))^2
}{ (1+e^\delta) (1+e^{\delta+u})}.
\]
(An approximate version of this formula was derived in \onlinecite{tsaousidou}.)
$1/ZT$ is plotted in Fig.~\ref{fig:ZTU} as a function of $u$ for
$\delta=2$ and $\gamma \rightarrow 0$. At $u=0$, the
non-interacting case, $ZT$ is divergent, as before. For $u \gg 1$,
$1/ZT$ goes to zero like $e^{-u}$. This can be understood within
the picture presented earlier. Even though, in contrast to the
non-interacting case there are now two levels (of energy
$\varepsilon_d$ and $\varepsilon_d+U$) on the molecule through which
transport can occur, the relative probability of electrons in the
leads having the corresponding energies goes as $e^{-u}$. As $U$
becomes large there is thus only an exponentially small
probability that there will be an electron in the leads with
energy $\varepsilon_d+U$. In this case, each electron that passes
through the junction carries heat $\varepsilon_d-\mu$ and the
situation is similar to the non-interacting case where transport
takes place through just one level on the molecule. Thus, the
thermal conductance in this case is also zero. (For convenience,
let us consider the symmetric case, with $\gamma_L=\gamma_R
=\gamma$.) Since the sequential tunneling model is correct only to
first order in $\gamma$, we can conclude that the thermal
conductance vanishes to first order in $\gamma$, and so must be of
order $\gamma^2$. For $\delta \simeq 2.4$ and $u\gg 1$, we have $G
S^2 \simeq \gamma 0.4k_B^2/\hbar$, and so $GS^2$ scales like
$\gamma$ as in the non-interacting case. Adding in once more the
phonon contribution to the heat current, we can again conclude
that $ZT$ will scale like $(\tilde{G}^{ph}_{th})^{-1/2}$.
In the limit of $u \rightarrow \infty$, the Green function method
\cite{meir_wingreen_1992,meir_wingreen_lee} can be used to find the
transport coefficients to higher order in $\gamma$, and so the
coefficient in the scaling of $ZT$.
The result is that, including the phonon contribution as before,
the optimal value of $ZT$ is $0.42 (G_{ph}^{th})^{-1/2}.$

%\section{Conclusion}
%\label{sec:conclusion}

We have shown that transport through a single molecular level can lead
to large values of the thermoelectric figure of merit $ZT$.
Within the Anderson model for the level, this occurs if
the molecular state is long-lived
(i.e., for the case of a symmetric junction,
$\gamma \rightarrow 0 $) and the energy
scale $U$ of the electron-electron
interactions within the molecule satisfies either
$U\ll k_BT$ or $U\gg k_BT$.
The increase in $ZT$ is due to a violation of the Wiedemann-Franz law ---
the charge conductance scales like $\gamma$, whereas the electron thermal
conductance divided by the temperature scales like $\gamma^2$.
%This violation is similar to that found in the work of Mahan and Sofo\cite{mahan_sofo} on bulk thermoelectric materials.
% They found that a material with a delta-function density of states for the conduction band has a similar anomaly in the Wiedemann-Franz law, and so can give large values of $ZT$.
While similarly large figures of merit may be achieved for optimal band structures in bulk materials~\cite{mahan_sofo}, the advantage of molecular systems is that they may have only a small phonon contribution to the thermal conductance, so that their
maximum achieved value of $ZT$ once phonons are included
may be higher than in bulk materials.

There are some practical challenges in achieving dramatically
enhanced $ZT$ in a real device by the mechanism presented here. The wavefunction of the level must be localized at the center of the molecule to give a small overlap with the states
in the leads, and so a long lifetime; in addition, the energy of
the level must lie a distance of about $2k_BT$ from the Fermi
level of the leads.  These requirements may be satisfied by the
Co(tpy-(CH$_2$)$_5$-SH)$_2$  molecules studied by Park et al
\cite{park_review,park_2002}. The electronic transport in these
systems is through a cobalt atom that sits at the center of the
molecule; the alkyl chains separate the cobalt from the gold
leads, creating a state localized away from the leads. It would be
of great interest to see if indeed the Wiedemann-Franz law is
violated in these molecules, as implied by the results in this paper.

\acknowledgments 

The authors thank A. Majumdar, P. Reddy, and R.
Segalman for useful discussions and acknowledge support from the
Division of Materials Sciences and Engineering of the Department
of Energy.

\bibliographystyle{./apsrev}
\bibliography{./ZTbib}
\end{document}